\documentclass{sigchi}

\usepackage{balance}  
\usepackage{graphics} 
\usepackage{times}    
\usepackage{url}      

\usepackage{ragged2e}
\usepackage{caption}
\usepackage{lipsum}
\usepackage{float}

\makeatletter
\def\url@leostyle{%
  \@ifundefined{selectfont}{\def\UrlFont{\sf}}{\def\UrlFont{\small\bf\ttfamily}}}
\makeatother
\def\pprw{8.5in}
\def\pprh{11in}

\usepackage[pdftex]{hyperref}
\hypersetup{
pdftitle={LDAExplore: Visualizing Topic Models Generated Using Latent Dirichlet Allocation},
pdfauthor={Ashwinkumar Ganesan, Kiante Branley, Jian Chen},
pdfkeywords={DataViz, InfoViz, LDA, Latent, Dirichlet Allocation, Parallel Coordinates, Treemap, Topic Modeling},
bookmarksnumbered,
pdfstartview={FitH},
colorlinks,
citecolor=black,
filecolor=black,
linkcolor=black,
urlcolor=black,
breaklinks=true,
}

\begin{document}

\title{LDAExplore: Visualizing Topic Models Generated Using Latent Dirichlet Allocation}

\numberofauthors{3}
\author{
  \alignauthor Ashwinkumar Ganesan, Kiante Brantley\\
    \affaddr{Dept. Of Computer Science \& Electrical Engineering, University Of Maryland Baltimore County}\\
    \email{\{gashwin1, bran4\}@umbc.edu}\\
  \alignauthor Shimei Pan\\
    \affaddr{Dept. Of Information Systems, University Of Maryland Baltimore County}\\
    \email{shimei@umbc.edu}\\
  \alignauthor Jian Chen\\
    \affaddr{Dept. Of Computer Science \& Electrical Engineering, University Of Maryland Baltimore County}\\
    \email{jichen@umbc.edu}
}

\maketitle

\begin{abstract}
We present \textit{LDAExplore}, a tool to visualize \textit{topic} distributions in a given document corpus that are generated using \textit{Topic Modeling} methods. Latent Dirichlet Allocation (LDA) is one of the basic methods that is predominantly used to generate \textit{topics}. One of the problems with methods like LDA is that users who apply them may not understand the \textit{topics} that are generated. Also, users may find it difficult to search correlated topics and correlated documents. \textit{LDAExplore}, tries to alleviate these problems by visualizing topic and word distributions generated from the document corpus and allowing the user to interact with them. The system is designed for users, who have minimal knowledge of \textit{LDA} or \textit{Topic Modelling} methods. To evaluate our design, we run a pilot study which uses the \textit{abstract}'s of 322 \textit{Information Visualization} papers, where every abstract is considered a \textit{document}. The topics generated are then explored by users. The results show that users are able to find correlated documents and group them based on topics that are similar.
\end{abstract}


\category{H.5.m.}{Information Interfaces and Presentation (e.g. HCI)}{Miscellaneous}

\section{Introduction}
\textit{Topic Modeling} tries to automate the process of extracting topics from documents while also annotate them with semantic information \cite{blei2012probabilistic}. It is made up of a set of statistical algorithms that extract correlated words from documents. These extracted word sets are called \textit{Topics}. Later, users can annotate the topic with semantic information. For example, consider a word set extracted which contains words such as \textit{visualization, sets, clusters, infoviz, interfaces} from the document corpus. Knowing the words associated with this \textit{topic}, we have a general notion that the word set represents the topic \textit{Information Visualization}. The ``topic'' generated is the word set and \textit{Information Visualization} is the semantic ``topic name'' annotated by the user. Latent Dirichlet Allocation (LDA) \cite{blei2003latent} is one of the common methods to perform topic modeling on a given corpus of documents.

LDA generates two types of distributions i.e. the topic distribution for each document in the set and the word distribution for each topic. These distributions can be changed by tweaking the hyper-parameters. \textit{LDAExplore}, tries to give visual cues about how these distributions look, and how the topics and documents are interrelated at the corpus level, between groups or individual documents. It is designed for users who may not know what topic modeling algorithms do. They are concerned with understanding the document corpus, finding the hidden topics in the corpus. {LDAExplore} gives the users the ability to search for documents using keywords from topics. 

One of the basic requirements of the design is that visual should scale for a large set of documents while providing the ability to see individual and group relations. In comparison to the set of documents, we assume that the number of topics is a much smaller set.

One of primary problems with topic modeling methods is that the ``topic'' generated by them, may not be clearly understood by the user \cite{chang2009reading}. One of the solutions to this problem, is to introduce a human-in-the-loop paradigm where users can interact with the algorithm, providing feedback such that the underlying model can be modified to generate ``better'' topics.

Our contributions in this paper are:
\begin{itemize}
\item Creating a comprehensive set of tasks required to design visualizations for text analysis using LDA.
\item A visualization that show correlations between topics and documents. The design can be scaled up to a larger set of documents without applying any aggregation method such as clustering to the documents.
\item Combining visual search \& filtrering to allow users to filter parallely across multiple topics in our design, so that users can easily filter a large document corpus.
\item A pilot study which shows the usability of each part of the tool.
\end{itemize}

\section{Related Work}
\label{sec:RelatedWork}

There are many different ways to visualize Topic Models. The methods include the use of force-directed graphs, parallel coordinates, matrices \& tree designs.

\subsection{Using Matrix Designs}
Matrix or tabular designs are easily understood by users. \textit{Termite} uses a tabular layout to promote comparison of terms both within and across latent topics. The primary visualization design used in it, is a matrix view where rows correspond to terms and columns to topics\cite{bertin1983semiology}. Another visualization \cite{chaney2012visualizing} uses a navigator to explore LDA generated topics. It shows the words in each topic and uses a tabular form to show which documents are associated. Once a specific document is selected the use can navigate to its Wikipedia article. These matrix layouts, show the correlation between terms and topics but have a difficulty scaling to a large number of documents. They do not show how the topic is correlated to the document, what its likelihood is.

\subsection{Using Trees \& Hierarchial Clustering}
\textit{RoseRiver} is another analytics system used to visualize how topics evolve \cite{cui2014hierarchical}. The system uses a tree cut approach with a combination of a word cloud. The word cloud is a standard method to show a word set with word sizes varying according to their frequency (or probability). Similarly, \textit{Overview} \cite{brehmer2014overview} is a technique designed for analyzing large text corpora using TF-IDF. It generates document clusters by hierarchically clustering these distances and encoding the result as a topic tree. \textit{VarifocalReader} \cite{koch2014varifocalreader} uses hierarchy structure such that it shows chapters, sub-chapters, and pages , and then lines of text. It again uses TF-IDF as one of its techniques. \textit{Serendip} \cite{alexanderserendip} is a technique designed for analyzing large text corpora using LDA. They present the results in a rectangle where each row means a document and each column means a topic.

Tree layouts can scale in design, but when combined with hierarchial clustering increase the amount of preprocessing that is necessary to present the tree structure. We choose not perform any clustering because user defined clusters and annotations, change the nature of the distributions generated by LDA.

\subsection{Using Force Directed Graphs \& Parallel Coordinates}
UTOPIAN \cite{choo2013utopian}, uses a force-directed graph to represent topics and is a semi-supervised system. It uses non-negative matrix factorization. Some other visualization include \textit{iVisClustering} \cite{lee2012ivisclustering} and \textit{ParallelTopics} \cite{dou2011paralleltopics}. Another parallel coordinate design which works with words in the topic directly is \textit{ThemeDelta}\cite{gadthemedelta}. Both have a design which uses parallel coordinates and \textit{iVisClustering} uses a force directed graph approach. Topic-based, interactive visual analysis tool (TIARA) \cite{pan2013optimizing} shows topic distributions across documents across time. Force-directed graphs are advantageous in terms of easy understanding \& usability, but they consume a lot of visual / screen space and may not scale for large number of terms or documents.

There are number of challenges \textit{LDAExplore} tries address i.e. a method that can scale for an increasing set of documents, a way to quantify the relation between documents and topics and provide and easy way to filter documents \& topics, while preserving their interrelations. In \textit{LDAExplore}, we try to reduce the amount of preprocessing necessary to display the topics, so that changes to the underlying model can be performed efficiently. We use parallel coordinates for showing document and topic relations while integrating keyword search to make it easy for the user to search for patterns using words from topics. This makes the filtering process more intuitive for the user.

\begin{figure*}[!t]
\centering
\includegraphics[scale=0.5]{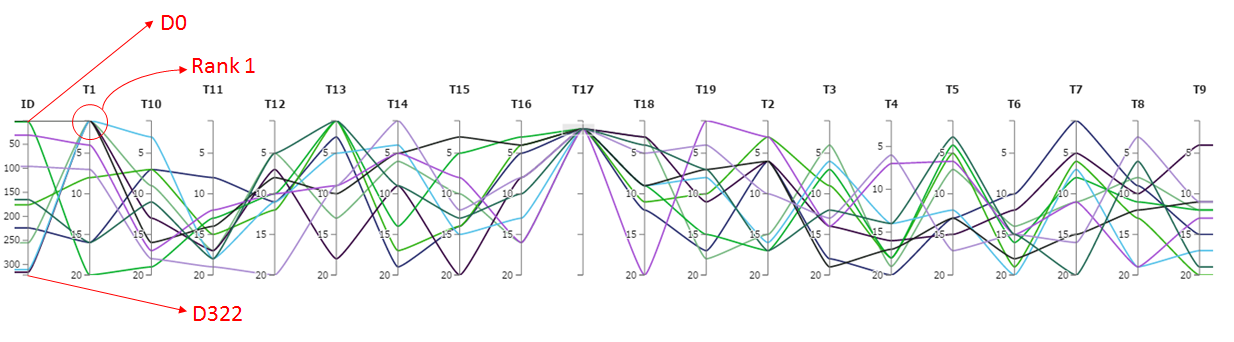}
\caption{\textbf{Displaying the Topic Distribution}}
\label{top_details}
\end{figure*}

\section{Design Considerations} 
\label{sec:design}

\subsection{Task Analysis}
As described in the previous section, results got from LDA, can be unintuitive. One of the main purpose of the visualization is to provide users with the options to explore the document set and give them the ability to provide feedback about topics to the system and which topics are correlated to documents, so that word and topic distributions are more intuitive and insightful. Following are the set of tasks which form the basis for our design.

\subsubsection{Visualize Topics}
LDA generates a set of topics, each having its own individual word list. Each word has a probability of being associated with the topic. As shown in figure \ref{top_words}, when the user interacts with a specific topic, the top words in the topic will be displayed, showing their probability. This is one way for the user to identify, what the topic is.

\subsubsection{Overview of Document - Topic Relations}
Once the user has a generic idea of the semantics of the topics, the user can then focus on the  correlations between documents and topics. The main purpose is to be able to see which topics are ranked higher over a large subset of documents, while which topics have a lower rank. The number of documents that rank a topic high, gives the user an idea as to which topics are more important.

\subsubsection{Remove Topics from the Visual}
When the user knows which topics are important, excess topics from the visual can be removed.

\subsubsection{Filtering Documents}
When the size of the corpus is large, the user will try to understand the distribution within a subset of the documents. Hence the user can filter the documents based on various criteria:
\begin{enumerate}
\item By top words in each document that associated with the topics having the highest probability of being related to the document.
\item The user may not know the documents, but may be able to identify some based on the \textit{name} or \textit{title} of the document. Thus, the user can filter out individual documents.
\end{enumerate}

\subsubsection{Perform set operations}
While working with a groups of documents, the user requires the ability to perform set operations such as \textit{Include} and \textit{Exclude}. \textit{Include} gives the user, the ability to add a filtered set of documents to future filters. \textit{Exclude} does the exact opposite, which is to remove the document set from any filter operations in the future. Once the document corpus has been explored, the user can export the filtered data for any post processing that is performed separately.

\subsubsection{Show \& Cluster Similar Topics}

Once the word distributions for each topic are known, the user will look for topics that are similar. Topic similarity can be measured using methods such KL-Divergence \cite{yangactive}. Similar topics can be grouped together, so that the number of topics on the visual display can be reduced and hierarchical collections of topics can be formed.

\subsubsection{Perform Cluster Operations}
Similar topics, will lead to a new group of documents which have these topics. The main task for the visual is to enable the user to define groups of documents and topics based on their own knowledge.

\subsubsection{Annotating Topics}
User should have the option to annotate the topics and documents or the respective clusters they lie in.

\begin{figure*}[!ht]
\centering
\includegraphics[scale=0.5]{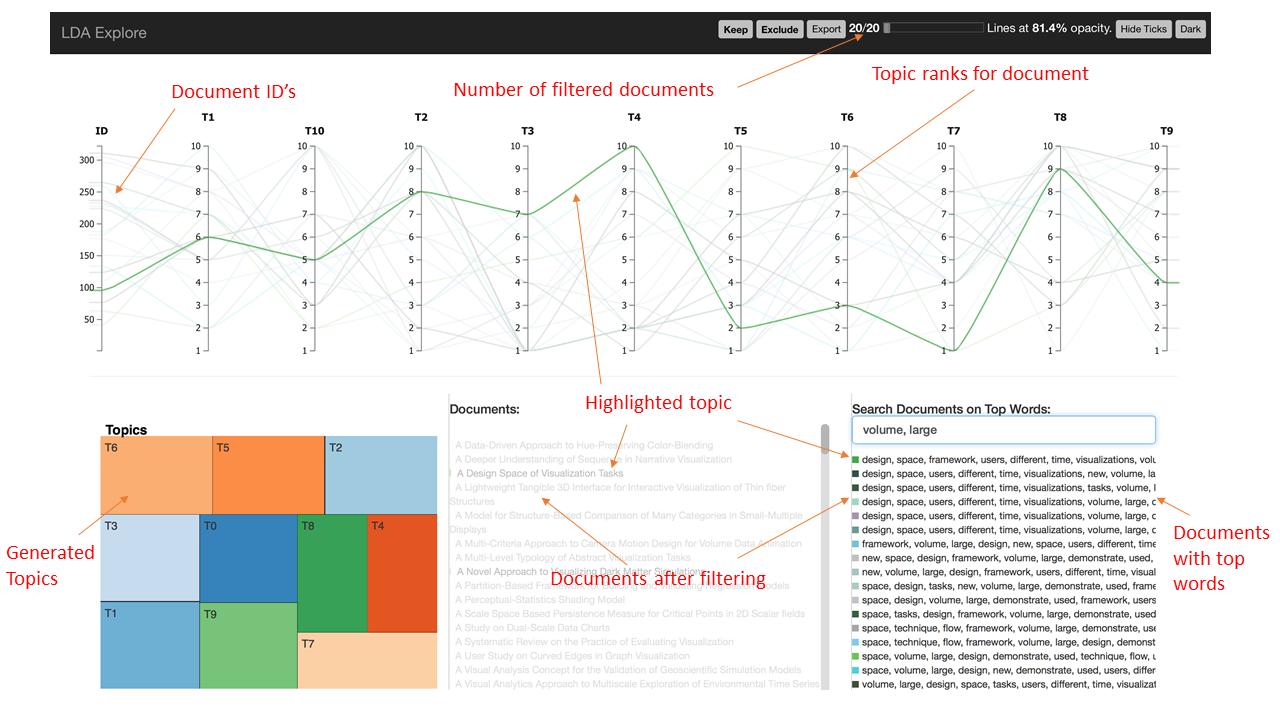}
\caption{\textbf{\textit{LDAExplore}: Filtering Mechanisms}}{Various sections of \textit{LDAExplore} are shown. The parallel coordinates shows the topic distribution, the treemap (which shows the word distribution for each topic on drill down), search tab used for filtering. A single filter is applied across the axis T17 to isolate documents ranked having T17 at a higher rank.}
\label{filtering}
\end{figure*}

\subsection{Prototype Design}
As described before, various sets, groups and LDA visualizations are available. They are able to display a very basic summarization of the document corpus and reveal basic relationships that occur. They provide the ability to explore individual documents, but they lack the ability to show document correlations across the dataset. When the correlations are shown, the design may not scale for a large number of documents. \textit{LDAExplore} creates a design to handle a sizeable number of documents while providing the capability to filter and visualize topics for individual documents. 

In its present form, \textit{LDAExplore} implements a subset of the tasks:
\begin{itemize}
\item Visualizing topics
\item Filtering \& Set operations
\item Displaying the overview of documents and topics correlations \item Removing topics. 
\end{itemize}

Based on the set of tasks defined, we describe how \textit{LDAExplore} works.

\subsubsection{Visualizing Topics}
We use a \textit{treemap} to visualize the topics. Figure \ref{filtering}, shows how the topics are displayed. Each rectangle is a topic and is given an ID like $T_1 ... T_n$ where there are $n$ defined topics. In figure \ref{top_details}, the total number of topics are $20$. In the treemap, the size of the rectangle defines the probability of the topic being associated to the document. In the current design, the likelihood is defined for each individual topic with respect to each document, rather than the document collection as a whole. Hence in figure \ref{filtering} the topics are represented by rectangles of equal area (but possibly different dimensions, depending on the number of topics) showing that all topics have an equal likelihood.

\begin{figure}[!h]
\centering
\includegraphics[scale=0.75]{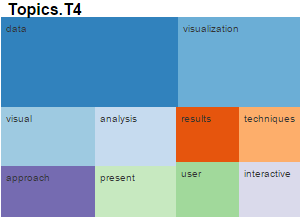}
\caption{\textbf{Top 10 words in Each Topic}}
\label{top_words}
\end{figure}

When the user clicks on a specific topic in the \textit{treemap}, the top $10$ words associated with the topic are revealed. Figure \ref{top_words}, shows an example word distribution. The word with the highest probability has the largest area. In this example, the word \textit{data} has the largest probability for topic \textit{T4}. In the current design, the number of words displayed per topic is maintained constant at $10$. The user can traverse back to the \textit{Topics}, by clicking on the \textit{Topics} tab at the top of the treemap. Whenever the user drills down, into the word distribution, the parent is always known.

\subsubsection{Relating Topics \& Documents}
We use parallel coordinates \cite{inselberg1991parallel}\cite{bostock2012data} to show the correlation between topics and documents. The parallel coordinates have two types of axes. The first one represents the document. The second type of axis is for topics. For example, consider Figure \ref{top_details}, which shows a total of $21$ axes. The first axis \textit{ID} shows the all the documents in the corpus and ranges from document ID $D_0$ to $D_{322}$. The next $20$ axes each represents a topic. Once LDA generates the probabilities of each topic in a document, we rank the topics. The topic axes shows the rank of the topic with respect to the document. So in figure \ref{top_details}, document $D_0$ has Topic $T_1$ at rank $1$, while others have it at rank $4, 7, 16 \& 20$ respectively. Each document has a different color, so that the user can distinguish between the graphs of various documents. The ranks are ordered in ascending order with $1$ showing the highest degree of correlation and $n$ (where n is the number of topics), showing $0$ or the least correlation. Topic ranks give the user which topic is the most correlated and which is the least.

In \textit{LDAExplore}, we decided to use parallel coordinates in order to be able to display the maximum amount of topics without the screen being too cluttered. When we tried to design other types of visualization, we quickly realized that scalability is an issue (by doing a quick claim analysis). In addition to parallel coordinate, we have a treemap to show the word in each topic. Being able see to the words in each topic is important because it allows the user to easily understand what that topic is about.

\subsubsection{Filtering}
\textit{LDAExplore} provides a range of filtering options. Figure \ref{filtering} provides a detailed overview of the filtering features. There are three main types of filtering:
\begin{enumerate}
\item \textit{Filter by Range} - Each axis can be used to filter the documents by selecting a specific range on the axis. The parallel coordinates displays the curves, only for the selected documents. Each axis can filtered simultaneously, thus creating a \textit{filter-chain}.
\item \textit{Filter by Searching} - A search bar is provided to the user, to search the documents using top $n$ words associated with each document. The current design, allows the user to search using any of the top $10$ words.
\item \textit{Filter by Selecting Individual Documents} - The documents column, lists each document in the corpus by its title. The user can click on a specific document to filter it out. Whenever a document is filtered out, its color is changed to downplay the document.
\end{enumerate}

Once the user has filtered the documents, the user can highlight a specific document within the filtered document set by using the mouse to hover over the document - word set (column 3 in figure \ref{filtering}). 
The total number of document in the filtered set is given at the top on the navigation bar. Also, the user can \textit{export} the filtered data to CSV format. 

\subsubsection{Set Operations}
There are $2$ main set operations that are supported in \textit{LDAExplore} i.e. include and exclude. The \textit{include} operation is performed by the \textit{Keep} and exclusion using \textit{Exclude} as seen in figure \ref{filtering}.

\subsubsection{Generating Top Words}
Once the documents are processed using LDA (from the gensim library \cite{vrehuuvrek2011gensim}), we end up with a list of topics and the probability of a word being associated with the topic.  There is a minimum probability threshold for each topic so that it can be considered in the final distribution. Topics below the minimum threshold are assigned a probability of 0. To generate the top words for each document we implement the following:

\begin{equation}
P(w_x,d_y) = \sum_{t=1}^{m} \sum_{i=1}^{n} P(w_x,t_i) \times P(t_i, d_y)
\label{word_prob}
\end{equation}

where $P(w_x,d_y)$ is the probability of a word with respect to a document, $P(w_x,t_i)$ is the probability of a word with respect to a topic and $P(t_i, d_y)$ is the probability of topic with respect to a document.

Equation \ref{word_prob} calculates the probability of each word in every topic for a specific document and then finds the top $k$ words which have the highest probability of being associated with the given document. This set of words generated are shown as the top $n$ searchable words on the visualization. Also, the words are used to create the topic-word hierarchy that is used in the treemap visualization for topics.

\subsubsection{Ranking Vs Probability}
The parallel coordinates can be used to display topic-document relations based on the rank of the topic (rank 1 being highest probable topic) or based on the actual probability of the topic. The current design displays the rank as it is easier to interpret by the user. As the difference probabilities between topics in a given distribution for a document reduce, it becomes harder for the users to define which topic are more important across a set of documents than the others on the parallel coordinates scale.

\subsubsection{Eliminate High Frequency Words}
The list of top words for each document and topic, contain many words which are common to most documents such as \textit{data} and \textit{visualization} as seen in figure \ref{filtering}. To alleviate this problem, Term Frequency - Inverse Document Frequency (\textit{TF-IDF}) \cite{rajaraman2011mining} is used, albeit not directly in combination with Latent Dirichlit Allocation (LDA). \textit{TF-IDF} is used to clear terms which are in high frequency across a large subset of the documents. It is executed with an arbitrary threshold of $t$, which describes the ratio of how many documents in the total corpus, can the word be present. Words over the threshold are from the final list of tokens before performing LDA.

\section{Pilot Study}
\subsection{Study Details}
The Pilot study is an initial study designed to gauge if the user can perform the tasks outlined, how usable the tool really is and get early feedback about the tool. It has been conducted using $5$ external participants. The responses from the first $3$ participants have been used to make some changes to the questions and then further feedback has been collected from the remaining $2$ participants. The number of participants having prior knowledge about \textit{Topic Modeling} and how LDA works is $2$. For the purpose of the study, the users have been asked to work with the tool on a standard wide screen monitor. The study has been partitioned into 4 parts i.e. the overview, topic visualization, filtering and keyword searching.

The questions in the study are of $4$ main types, Tasks execution questions (such as \textit{How many documents are represented in this visual?}), Understanding questions (such as \textit{Does the visual have too many things on it?}), Reasoning questions (such as \textit{Which is the least important topic? (Important means highly ranked topic)}) and Usability questions (such as \textit{Are the rankings on the axis visible / readable?}).

We provide details on how to access the tool, a brief overview of the features of the tool with the help of an annotated example diagram (such as figure \ref{filtering}). Also, the users are given information about the data set, they will be visualizing. For the purpose of our study, the data set consists of the abstracts from $322$ research papers from the domain of \textit{Data Visualization}. Each abstract is considered as a separate document. The information contains attributes like the title of the paper and the abstract of the research paper. LDA is used to identify the plausible underlying topic / topics for each research paper. The \textit{Title} of paper that is seen in the ``Documents'' column and the abstract of each research paper. The next section describes each part of the study and the associated results.

\subsection{Survey \& Results}
\subsubsection{Overview}
The overview questions are designed to understand if participants can understand any useful information immediately after loading the main page and check what they think about the overall look and feel of the tool. The main page shows all the data without any kind of filtering.

The users find the initially displayed parallel coordinate to be slightly cluttered, when it is presented without any filtering. Also, the users find it difficult to search for markings on the axis in the parallel coordinate such as axis legends. The overall information on the number of topics and the number of documents represented in the visual can be deduced easily.

\subsubsection{Visualizing Topics}
The \textit{topics} questions are designed to see if the participants can understand what kind of topics are generated by LDA and what are the words (with their individual probabilities) associated with the topic. This section is very important because while inspecting LDA results, users are likely to investigate what topics are produced from a corpus of documents.

The users are able to understand which topics have which words and which of them are important within these topics. The results confirm that the treemap is useful in visualizing topics and that the areas of rectangles used to define which topic / word is more or less important (larger size being more important) is useful.

\subsubsection{Filtering}
Filtering questions are designed to see if users can use all the features within \textit{LDAExplore} to effectively filter documents to find specific patterns. Users are tasked with filtering a specific topic's axis on the parallel coordinate for a range of ranks and isolate the subset of documents which find the topic to be important. Users are then asked to find topics that might be similar and documents (research papers) that can be grouped together. The study shows users are able to filter the document corpus easily. Some users find it difficult to deduce which topic is more or less important. This because the parallel coordinates design has a single line for each document and does not \textit{bundle} lines together at every rank of each topic, thus giving the user an idea of the number of documents at each rank.

\subsubsection{Keyword Search}

The last set of questions are on the usefulness of \textit{keyword} search. Users are able to see the effect of \textit{filtering using keywords} on other sections (documents \& search) of the tool. Searching is one of more useful features of the tool because it instantaneously reduces the number of lines in the parallel coordinate and provides immediate information about documents based on the user's search query.

\subsection{Recommended Changes to the Design}

Based on the information gathered from the study, we have a list of changes that can be done. To reduce the clutter in the parallel coordinates (for large corpora), edge bundling can be introduced \cite{holten2006hierarchical}. To de-clutter it further, documents can be clustered to visualize the topics that are ranked for these groups. The topics themselves can be grouped together based upon their similarity. The navigation between topics in the treemap can be improved. 

Users in the study find it diffcult to understand the ``Documents'' column. The column has titles of documents and acts as filter to remove a document from the parallel coordinate rather than selectively show it. This is counter intuitive to how humans use \textit{URL's/ clicks} on an item,  which is mainly to select a page or the highlighted item. Hence this filter can be removed or altered to select a specific document and filter out the rest of the corpus (or highlight the specific document with respect to others).

The treemap (topics) and parallel coordinate(topic distribution) are disconnected. This can be modified, so that the treemap can be used as additional filtering where the documents across the parallel coordinates can be filtered using topic words. To find out which topics have similar probability in the topic distribution for a given document, user can be given the option to switch between the rank mode and probabilistic modes in the visualization.

Searching using words is useful, but there is not mechanism to search the corpus for a specific document using attributes such as the title. Searching can be improved by adding the title of the document to the \textit{keywords search} section.

Once the documents have been filtered users need a way to \textit{save} the selection, so that it can be utilized at a later time. This \textit{in-memory} storage of the filter applied can be useful.

\section{Conclusion}

Being able to understand LDA results through a visualization is a critical part of helping end users to explore documents and recognize patterns. This paper describes \textit{LDAExplore}, a tool to visualize a document corpus. It provides the results of a pilot study that shows that the targeted audience have a better understanding of what the corpus contains \& what LDA results mean.

We have worked closely with $5$ individuals, $2$ of whom are able to understand LDA. They are able to navigate through the visualization with minimal guidance in order to solve the tasks such as finding documents which are related to specific topics that are difficult to do without a visualization.

In the future we plan to investigate how to enable users to cluster topics if they decide that the topics selected are not distinct enough. Similarly, users will like to cluster documents whose topics look similar and see how these aggregated clusters differ. User defined clustering, changes the underlying model within LDA. It will be interesting to see how the changes in model affect other topics or documents.

\section{Acknowledgments}
We would like to thank Li-Chien Lee for the help on this paper and all the participants of the study, who spent a portion of their valuable time evaluating our project.

This work was supported in part by grants from NSF IIS-1302755, IIS-1341254, DBI-1260795,  DBI-1341352, and NIST 70NANB13H181. Any opinions, findings, and conclusions or recommendations expressed in this material are those of the author(s) and do not necessarily reflect the views of the National Science Foundation. Certain commercial products are identified in this paper in order to specify the experimental procedure adequately. Such identification is not intended to imply recommendation or endorsement by the National Institute of Standards and Technology, nor is it intended to imply that the products identified are necessarily the best available for the purpose.

\balance
\bibliographystyle{acm-sigchi}
\bibliography{main}
\end{document}